\documentclass{article}
\usepackage{spconf}
\ninept
\usepackage{tikz}
\usepackage{amsmath,amsfonts,amssymb}%
\usepackage{bm}%
\usepackage{graphicx,graphics}%
\usepackage{cases}%
\usepackage[noadjust]{cite}%
\usepackage{color}%
\usepackage{cite,url}
\usepackage{verbatim}
\usepackage{algorithm}
\usepackage{balance}
\usepackage{multirow}
\usepackage{stfloats}
\usepackage{xspace}
\usepackage{amsthm}
\usepackage{mathtools} % \DeclareDelimiter, smashoperator, etc
\usepackage{subfig} % subfloat

\usepackage{caption}

\usepackage{tikz}
\usetikzlibrary{shapes.misc}
\usetikzlibrary{matrix}
\usetikzlibrary{arrows,backgrounds,fit,calc}

\usepackage{float}
\floatname{algorithm}{Procedure}

\newcommand{\bv}[1]{\boldsymbol{#1}}
\newcommand{\dfn}{\triangleq}
\newcommand{\untsph}{\mathbb{S}^{2}} % unit sphere

\newcommand{\shc}[3]{({#1})_{#2}^{#3}}
\newcommand{\lsph}{L^2(\untsph)}
\newcommand{\unit}[1]{\bv{\hat{#1}}}
\newcommand{\lsphL}[1]{\mathcal{H}_{#1}}

\newcommand{\intsph}{\int_{\untsph}}

\newcommand{\figref}[1]{Fig.\,\ref{#1}}

\DeclarePairedDelimiterX\innerp[2]{\langle}{\rangle}{#1,#2}

\graphicspath{{figs/},{pdfs/},{pdf/},{Figures/},{Authors/}}

\begin{document}
\title{Estimation of Groundwater Storage Variations in Indus River Basin using GRACE Data}

	\name{Yahya Sattar$^{\dag}$, {Student Member,~IEEE} and  Zubair~Khalid$^{\star}$, {Senior Member,~IEEE}}

\address{  $^{\dagger}$ Department of Electrical and Computer Engineering, University of California, Riverside, CA 92521
\\  $^{\star}$ School of Science and Engineering, Lahore University of Management Sciences, Lahore, Pakistan\\
Email: ysattar@engr.ucr.edu, zubair.khalid@lums.edu.pk}

\maketitle

\begin{abstract}
	The depletion and variations of groundwater storage~(GWS) are of critical importance for sustainable groundwater management. In this work, we use Gravity Recovery and Climate Experiment (GRACE) to estimate variations in the terrestrial water storage~(TWS) and use it in conjunction with the Global Land Data Assimilation System~(GLDAS) data to extract GWS variations over time for Indus river basin~(IRB). We present a data processing framework that processes and combines these data-sets to provide an estimate of GWS changes. We also present the design of a band-limited optimally concentrated window function for spatial localization of the data in the region of interest. We construct the so-called optimal window for the IRB region and use it in our processing framework to analyze the GWS variations from $2005$ to $2015$. Our analysis reveals the expected seasonal variations in GWS and signifies groundwater depletion on average over the time period. Our proposed processing framework can be used to analyze spatio-temporal variations in TWS and GWS for any region of interest.
\end{abstract}

\begin{keywords}
	GRACE, groundwater storage, GLDAS, 2-sphere, terrestrial water storage, window design, spherical harmonics.
\end{keywords}

% \fi

\section{Introduction}

The primary source of water for more than 2 billion people is groundwater and its excessive withdrawal calls for the monitoring of groundwater to sustain the expansion of the human population~\cite{Famiglietti:2011}. The conventional groundwater monitoring methods are not only time and money consuming, but also limited by their spatial coverage, and therefore cannot produce accurate dynamic observations over the large spatial region~\cite{Longuevergne:2013}. The monitoring of groundwater is even more challenging for the case of the Indus river basin (IRB) where around $60\%$ of the water is being pumped from the groundwater with a lack of planning and control on the pumping and drilling of wells~\cite{Naomi:2019}. In this work, we propose to process NASA's Gravity Recovery and Climate Experiment~(GRACE) satellite mission data to monitor dynamic changes in groundwater storage in the Indus River Basin (IRB).

\subsection{Relation to Prior Work}

GRACE data has been previously used in conjunction with data obtained from hydrological models to monitor groundwater storage changes in large areas or river basins on a seasonal or annual scale~(e.g., \cite{Swenson:2002,Yeh:2006,Tiwari:2009,Rodell:2009,Famiglietti:2011,Longuevergne:2013,Jiang:2014,Xiao:2015}). Various methods have been used to estimate the variations in groundwater storage~(GWS) in large regions. An extensive method has been employed in~\cite{Rodell:2009} for monitoring GWS depletion in India over a period of 6 years concluding that the {groundwater is being depleted at a mean rate of $4.0 \pm 1.0 {\rm cm/yr}$ equivalent height of water}~($17.7 \pm 4.5 {\rm km^3/yr}$ in volume) around the Indian states of Rajasthan, Punjab and Huryana~(including Delhi). In the existing literature, the proposed methods use GRACE data in conjunction with the hydrological model to estimate terrestrial water storage~(TWS) changes. For obtaining localized estimates, spatial averaging is used over the spatial region on the sphere. Since this is equivalent to boxcar windowing of the data, the localized estimates are not accurate due to the infinite spherical harmonic band-limit~(formally defined in Section 2.2) of the spatially localized signal~\cite{Wieczorek:2005}. In these methods, the change in GWS is computed by estimating soil-water storage variations using Global Land Data Assimilation System~(GLDAS) followed by removing it from the change in TWS. Employing a similar processing of the GRACE data, the work in \cite{Tiwari:2009} used Community Land Model~(CLM) which includes both groundwater and a river storage component, to subtract soil-water~(lakes, reservoirs, glaciers and soil moisture, etc.) variations from GRACE TWS variations to obtain an estimate of GWS changes in the regions of Northern India, which revealed that the region lost groundwater at a rate of $54 \pm 9 {\rm km^3/yr}$ in volume between April $2002$ and June $2008$. Groundwater depletion has also been estimated in the Central Valley of California in~\cite{Famiglietti:2011}, where 78 months of data from the GRACE satellite mission is used to estimate water storage changes in Sacramento and San Joaquin River Basins indicating that the Central Valley lost $20.4 \pm 3.9 {\rm mm/yr}$ of groundwater during the 78-months period.

\subsection{Contributions}

We use GRACE data in conjunction with the hydrological models and GLDAS data to determine spatio-temporal variations in GWS in the Indus river basin (IRB). We first compute TWS using GRACE data and remove snow water equivalent and soil moisture storage contributions from it to obtain GWS variations. In our data processing framework, we use a window function to obtain localization of the signal in the region of interest. We present the design of an optimal window function for spatial localization in the IRB region. The window function is optimal in the sense that it is band-limited in the spherical harmonic domain and maximally concentrated in the spatial region~(IRB). We estimate variations in GWS for IRB from $2005$ to $2015$ using our framework and demonstrate the temporal changes in the GWS in addition to the depletion of GWS along time. We organize the rest of the paper as follows. We review the mathematical preliminaries in Section 2 before presenting the proposed processing framework, window design and results in Section 3 and making concluding remarks in Section 4.

\section{Mathematical Background and Preliminaries} \label{sec:preminaries}

\subsection{Signals on the Sphere}\label{subsec:sphere}

A point $\unit{u}\in\mathbb{R}^3$ on the unit sphere~(or 2 sphere), denoted by $\mathbb{S}^2$, can be parameterized in terms of two angles namely colatitude $\theta \in [0,\,\pi]$ and longitude $\phi \in [0, \, 2\pi)$ as $\unit{u} \equiv \unit{u}(\theta,\phi) \triangleq (\sin\theta\,\cos\phi, \; \sin\theta\,\sin\phi, \; \cos\theta)$. The angle colatitude is measured from the positive $z$-axis and the longitude is measured from the positive $x$-axis in the $x$-$y$ plane. The complex valued and square-integrable functions~(or signals) defined on the unit sphere form a Hilbert space $\lsph$ equipped with the following inner product for two functions $g, h\in\lsph$:
\begin{align}\label{eqn:innprd}
\langle g, h \rangle \triangleq  \int_{\mathbb{S}^2} g(\unit{u}) \,\overline {h(\unit{u})} \,ds(\unit{u}),
\end{align}
where $\overline{(\cdot)}$ denotes the complex conjugate, $ds(\unit{u}) = \sin\theta\, d\theta\,d\phi$ is the differential area element on $\mathbb{S}^2$ and $\int_{\mathbb{S}^2} = \int_{\theta=0}^\pi  \int_{\phi=0}^{2\pi}$.

\subsection{Spherical Harmonics}

Spherical harmonic functions are denoted by $Y_{\ell}^{m}(\unit{u})\equiv Y_{\ell}^{m}(\theta, \phi)$ for integer degree $\ell \geq 0$ and integer order $|m| \leq \ell$~\cite{Kennedy-book:2013}. Spherical harmonics serve as complete orthonormal basis functions for the Hilbert space $\lsph$ and therefore we can expand any signal $g\in\lsph$ using spherical harmonics as
\vspace{-1mm}
\begin{align}\label{eqn:HarmonExpan}
g(\unit{u}) = \sum\limits_{\ell,m}^{\infty} \shc{g}{\ell m}{} \, Y_{\ell}^m(\unit{u}), \quad \sum\limits_{\ell,m}^{\infty} \equiv \sum\limits_{\ell=0}^{\infty} \sum\limits_{m=-\ell}^{\ell},
\end{align}
where $\shc{g}{\ell m}{} \dfn \langle g, Y_{\ell}^m \rangle$ is the spherical harmonic coefficient of degree $\ell \geq 0$ and integer order $|m| \leq \ell$. The spherical harmonic coefficients represent the signal in harmonic~(Fourier) domain. We refer to the signal $g$ bandlimited at degree $L$ if $\shc{g}{\ell m}{}=0,\,\forall\, \ell \ge L,\, -\ell \leq m \leq \ell$. Set of all such bandlimited signals on the sphere forms an $L^2$-dimensional subspace $\lsphL{L}$ of $\lsph$ and their spherical harmonic coefficients can be stored in an $L^2 \times 1$ column vector as
\begin{align}
\bv{g} = [(g)_{0,0},(g)_{1,-1},(g)_{1,0},(g)_{1,1},\ldots,(g)_{L-1,L-1}]^{\mathrm{T}}.
\label{eq:flm_column}
\end{align}

\subsection{GRACE and GLDAS Data Sets}

We use Gravity Recovery and Climate Experiment~(GRACE)\footnote{
GRACE mission was jointly launched by {NASA} and {DLR}~(the German Aerospace Center) in March, 2002 with an objective to provide spatio-temporal variations in the gravity field in the form of Stokes coefficients with monthly temporal resolution~\cite{Grace:2010}.} gravity solutions\footnote{\url{https://podaac.jpl.nasa.gov/dataset/GRACE_L1B_GRAV_JPL_RL03}}~(Release 03). This dataset consists of a set of Stokes coefficients $\overline{C}_{\ell m}$ and $\overline{S}_{\ell m}$ both for integer degree $0 \leq \ell$ and integer order $0 \leq m \leq \ell$. The Stokes coefficients are used to compute~(monthly) surface density change $\Delta\sigma$ over the unit-sphere as follows~\cite{Wahr:1998},
\begin{equation}\label{eqn:deltaOmega}
\begin{aligned}
	\Delta\sigma(\theta, \phi) = \frac{a \rho_a}{3} \sum\limits_{\ell, m}^{\infty} \frac{2\ell + 1}{1 + k_{\ell}^{\prime}}\big(\Delta\overline{C}_{\ell m}\,X_{\ell}^m(\theta)\cos\theta  \\ +\Delta\overline{S}_{\ell m}\,X_{\ell}^m(\theta)\sin\theta\big),
\end{aligned}
\end{equation}
where $a = 6371.008$ km is the average radius of the earth, $\rho_a = 5517$ ${\rm kg/m^3}$ is the average density of the earth, $k_{\ell}^{\prime}$ is the load Love number at degree $\ell$ and $X_{\ell}^m(\theta)$ is the associated Legendre function~\cite{Kennedy-book:2013}. The surface density change is proportional to the variations in total water storage~(TWS) in the units of equivalent water height. The major causes of TWS variations over time include i) precipitation stored as snow, ii) water penetrated into the ground, iii) water evaporated or departed the basin as stream flow and iv) water pumped out from natural undergroundwater reservoirs.

We also use the Global Land Data Assimilation System~(GLDAS) to obtain the estimates of the soil moisture changes, snow water changes, and changes in the surface water. GLDAS is available in the spatial domain with equiangular resolution of $\frac{1}{4^\circ}$ and uses the advance land surface modeling and data assimilation techniques, in order to generate optimal fields of land surface states~(e.g., soil moisture) and fluxes~(e.g., evapotranspiration)~\cite{Rodell:2004}.

\section{Estimation of groundwater Storage Variations}\label{sec:main}

\subsection{Proposed Framework Overview}
We estimate the variations in GWS in the Indus river basin~(IRB) by processing the GRACE data from $2005$ to $2015$. We first remove the temporal mean from the Stokes coefficients $\overline{C}_{\ell m}^{(t)}$ and $\overline{S}_{\ell m}^{(t)}$ to obtain gravitational anomalies coefficients $\Delta\overline{C}_{\ell m}^{(t)}$ and $\Delta\overline{S}_{\ell m}^{(t)}$. We band-limit the Stokes coefficients at spherical harmonic degree $L$ and use these coefficients to obtain spherical harmonic coefficients as
\begin{align}
	\Delta\shc{f}{\ell m}{(t)} = \Delta\overline{C}_{\ell m}^{(t)} - i\Delta\overline{S}_{\ell m}^{(t)}
\end{align}
for all non-negative orders $m$. Noting that the gravitational anomaly signal $f$ is real-valued, we obtain the spherical harmonic coefficients by employing the conjugate symmetry relationship given by $\Delta\shc{f}{\ell,-m}{(t)} = (-1)^m \Delta\overline{\shc{f}{\ell m}{(t)}}$. We also carry out smoothing of the signal by convolving it with the von Mises function~(spherical analogue of Gaussian smoothing) defined as
\[
h(\theta,\phi)=h(\theta) = 	\frac{\kappa\exp(\kappa\,\cos\theta)}{4\pi\sinh\kappa}.
\]
The smoothing is carried out in the harmonic domain to obtain the spherical harmonic coefficients of the filtered signal $\mathcal{G}$ as~\cite{khalid2014choice}
\begin{align}
\shc{\mathcal{G}}{\ell m}{(t)} = {\frac{\mathcal{I}_{\ell+1/2}(\kappa)}{\mathcal{I}_{1/2}(\kappa)}} \Delta\shc{f}{\ell m}{(t)},
\end{align}
where $\mathcal{I}_{\ell+1/2}$  is a half-integer-order modified Bessel function
of the first kind and $\kappa>0$ is a constant that controls the degree of smoothness. Using the filtered coefficients, we determine the temporal change in surface density of the Earth as~\cite{Wahr:1998}
\begin{equation}\label{eqn:delta_sigma}
	\begin{aligned}
		\Delta \sigma^{(t)}(\theta, \phi) = \frac{a \rho_a}{3}\sum\limits_{\ell,m}^{L-1} \frac{2\ell + 1}{1 + k_{\ell}^{\prime}}\shc{\mathcal{G}}{\ell m}{(t)}Y_{\ell}^m(\theta, \phi),
	\end{aligned}
\end{equation}
where $a = 6371.008$ km is the average radius of the earth, $\rho_a = 5517$ ${\rm kg/m^3}$ is the average density of the earth and $k_{\ell}^{\prime}$ is the load Love number at degree $\ell$ that measures the rigidity of a planetary body and the susceptibility of its shape to change in response to a tidal potential. Normalizing $\Delta\sigma^{(t)}(\theta, \phi)$ by the water density $\rho_w = 1000$ ${\rm kg/m^3}$ yields spatio-temporal variations in the terrestrial water storage~(TWS), that is,
\begin{align}
	\Delta {\rm TWS}^{(t)}(\theta,\phi) = (1/\rho_w)\Delta \sigma^{(t)}(\theta, \phi). \label{eqn: deltaTWS_globe}
\end{align}

In order to obtain estimate of TWS over a region~(e.g., IRB), we need to average the localized estimate of variations in the terrestrial water storage. The use of boxcar window for the localization of $\Delta {\rm TWS}^{(t)}(\theta,\phi)$, given in \eqref{eqn: deltaTWS_globe}, inside the region $R$ yields infinite band-limit signal, the average of which cannot be computed by taking finite number of samples over the region.

\subsection{Optimal Window Design for Localization}

To localize a global signal over the region of interest, we propose to use band-limited window function with optimal energy concentration in the spatial region. For spatial localization of TWS in IRB, we design window function by solving spherical Slepian spatial-spectral concentration problem~\cite{Simons:2006} which seeks to maximize the energy concentration ratio of a band-limited signal $f\in \lsphL{L}$ within the spatial region $R$, that is,
\begin{align}
\lambda &=  \frac{\int_R |f(\unit{u})|^2 ds(\unit{u})} {\intsph |f(\unit{u})|^2ds(\unit{u})}
 =  \frac{ \sum\limits_{\ell,m}^{L-1} \sum\limits_{p,q}^{L-1} \overline{(f)_{\ell}^m} (f)_p^q K_{\ell m,p q} }{\sum\limits_{\ell,m}^{L-1} |(f)_{\ell}^m|^2} &= \frac{ \bv{f}^{\mathrm{H}} \bv{K} \bv{f} }{\bv{f}^{\mathrm{H}} \bv{f}},
\label{eq:e_ratio}
\end{align}
where $|f(\unit{u})|^2 =  f(\unit{u})\,\overline{f(\unit{u})}$, we have used the orthonormality of spherical harmonics on the sphere to obtain the second equality, $(\cdot)^{\mathrm{H}}$ denotes conjugate transpose operation and the indexing introduced in \eqref{eq:flm_column} is adopted to define $L^2\times L^2$ matrix $\bv{K}$ with entries
\begin{align}
K_{\ell m,p q} \dfn \int_R \overline{Y_{\ell}^m(\theta,\phi)} Y_p^q(\theta,\phi) \, \sin\theta \, d\theta d\phi.
\label{eq:K_elements}
\end{align}
Maximization of the energy concentration ratio, formulated in \eqref{eq:e_ratio} is equivalent to finding the solution of the following eigenvalue problem:
\begin{align}
\bv{K} \bv{f} = \lambda \bv{f},
\label{eq:ev_problem}
\end{align}
the solution of which provides $L^2$ band-limited eigenfunctions on the sphere and the eigenvalue $\lambda$ quantifies the energy concentration of the associated eigenfunction within the region $R$. We propose to use eigenfunction with maximum concentration~(largest eigenvalue) in the region of interest for localization of the signal and refer to it as an optimal window function. We consider to design a window function as a weighted sum of eigenfunctions with near optimal concentration in the region of interest as future research direction.

\begin{figure}[t]
	\centering
	%\vspace{-0.8in}
	\includegraphics[scale=.4]{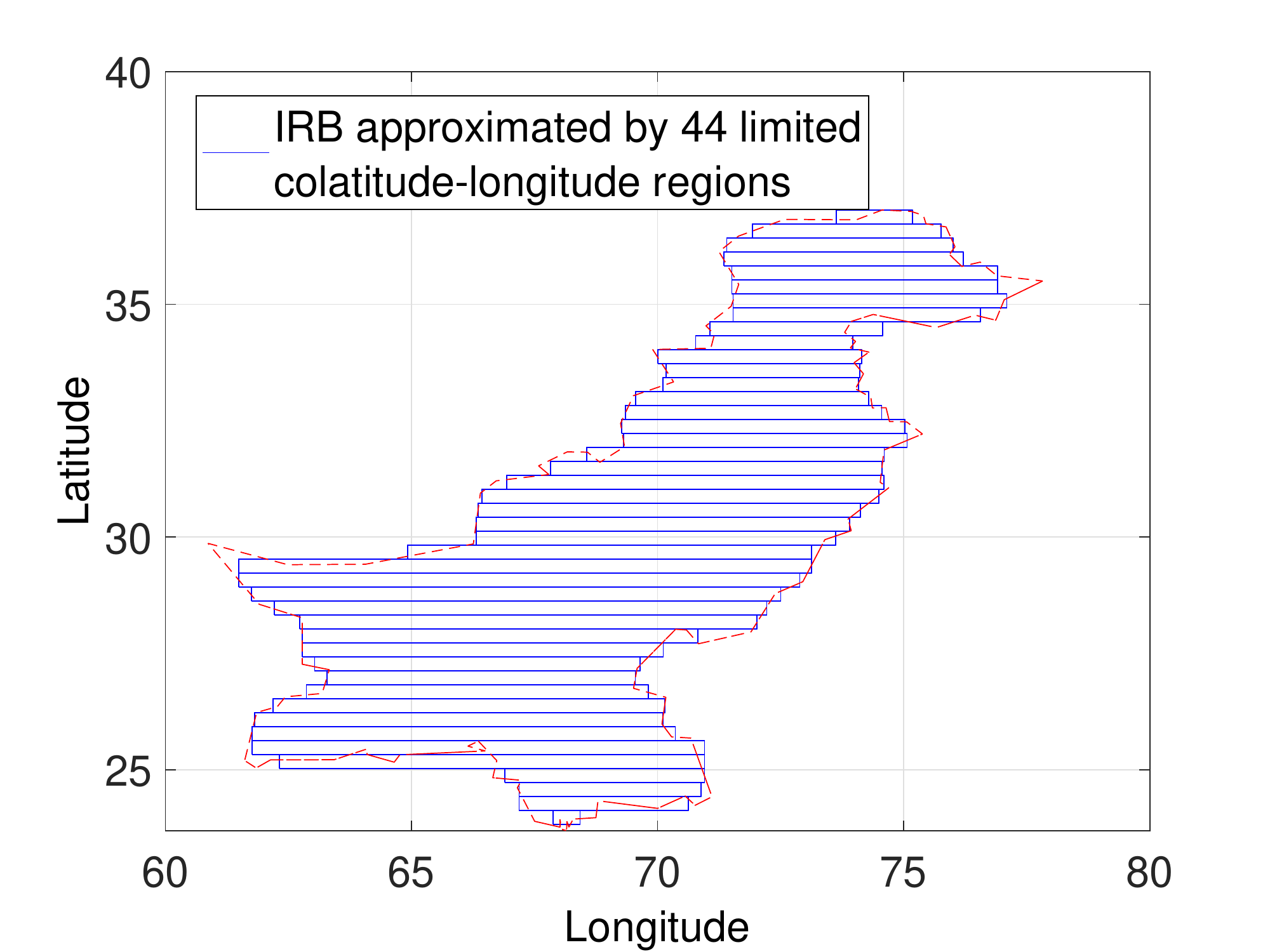}
	%\vspace{-2.5in}
	\caption{We approximated the IRB by $44$ limited colatitude-longitude regions for analytic computation of the integral in  \eqref{eq:K_elements2} which is required to design optimal window function for spatial localization.}
	\label{fig:pakistan_region}
\end{figure}

In order to compute $K_{\ell m,p q}$ for IRB region $R$, we divide it into 44 limited colatitude-longitude subregions, as elaborated in \figref{fig:pakistan_region}, such that each subregion $\tilde{R}_k$ can be parameterized as
\begin{align*}
	\tilde{R}_k \triangleq \{(\theta,\phi): \theta_{k,1} \leq \theta \leq \theta_{k,2}, \; \phi_{k,1} \leq \phi \leq \phi_{k,2} \}.
\end{align*}
We compute the integral in \eqref{eq:K_elements} by evaluating it over each subregion. For each subregion $\tilde{R}_k$, we define \begin{align}
K^k_{\ell m,p q} = \int_{\tilde{R}_k} \overline{Y_{\ell}^m(\theta,\phi)} Y_p^q(\theta,\phi) \, \sin\theta \, d\theta d\phi,
\label{eq:K_elements2}
\end{align}
which can be evaluated using the following analytical expression~\cite{Alice:2016}
\begin{align*}
	K_{\ell m, pq}^k = \sum\limits_{m' = -\ell}^{\ell} \sum\limits_{q' = -p}^p F_{m', m}^{\ell} F_{q', q}^p Q(m' + q') S(q - m),
\end{align*}
where $Q(m)$, $S(m)$ and $F_{m', m}^{\ell}$ are evaluated as
\begin{align*}
	Q(m) &=
	\begin{cases}
		\frac{1}{4}(i2m(\theta_{k,2} - \theta_{k,1}) + e^{i2m\theta_{k,1}} - e^{i2m\theta_{k,2}}),~|m| = 1 \\
		\frac{1}{m^2 -1}(e^{im\theta_{k,1}}(-\cos\theta_{k,1} + im\sin\theta_{k,1}) \\
		\quad~~+(e^{im\theta_{k,2}}(\cos\theta_{k,2} - im\sin\theta_{k,2})), ~~~~~~~~|m| \neq 1
	\end{cases}\\
	S(m) &=
	\begin{cases}
	 \phi_{k,2} - \phi_{k,1}, \quad\quad\quad\quad\quad~~ m=0 \\
	\frac{i}{m}(e^{im \phi_{k,1}}-e^{im \phi_{k,2}}), \quad m \neq 0
	\end{cases}\\
F_{m', m}^{\ell} &= (-i)^m \sqrt{\frac{2\ell + 1}{4\pi}} \Delta_{m',m}^{\ell} \Delta_{m,0}^{\ell},
\end{align*}
%\begin{align}
%	F_{m', m}^{\ell} = (-i)^m \sqrt{\frac{2\ell + 1}{4\pi}} \Delta_{m',m}^{\ell} \Delta_{m,0}^{\ell},
%\end{align}
where $\Delta_{m,n}^{\ell} \triangleq d_{m,n}^{\ell}(\pi/2)$ is the Wigner-$d$ function~\cite{Kennedy-book:2013}. %We plot the window function used for localization in \zk{Figure to be added}

\subsection{Estimation of Groundwater Storage}
Using the proposed optimal window, designed for IRB and denoted by $\mathcal{W}_{\rm IRB}$, we first determine the variations in TWS over the Indus river basin as
\begin{align}
\Delta {\rm TWS}^{(t)}_{\rm IRB}(\theta,\phi) = \mathcal{W}_{\rm IRB}(\theta, \phi)\Delta {\rm TWS}^{(t)}(\theta,\phi). \label{eqn: deltaTWS_IRB}
\end{align}
Next, we use GLDAS data to to estimate changes in snow water equivalent~($\Delta{\rm SWE}$) and soil moisture storage~($\Delta{\rm SMS}$) over time. Since the GLDAS data~(denoted by $g$) is available in the spatial domain, we compute spherical harmonic transform to obtain its spherical harmonic~(SH) coefficients.
\begin{align}
	\big(g\big)_{\ell m}^{(t)} = \langle g^{(t)}(\theta,\phi), Y_{\ell m}(\theta, \phi) \rangle. \label{eqn: GLDAS SH}
\end{align}
We then remove the temporal mean to obtain $\Delta\big(g\big)_{\ell m}^{(t)}$, band-limit the spherical harmonic coefficients at degree $L$ and apply Gaussian smoothing. After processing in the harmonic domain, the variations in SWE or SMS is computed by taking the inverse SH transform as follows:
\begin{align}
\Delta g^{(t)}(\theta,\phi) = \sum\limits_{\ell,m}^{L-1} {\frac{\mathcal{I}_{\ell+1/2}(\kappa)}{\mathcal{I}_{1/2}(\kappa)}}\Delta\shc{g}{\ell m}{(t)} Y_{\ell}^m(\theta, \phi). \label{eqn: GLDAS var}
\end{align}
Finally, the localization is carried using the proposed optimal window to obtain variations  in SWE or SMS in the IRB, that is,
\begin{align}
\Delta g^{(t)}_{\rm IRB}(\theta,\phi) = \mathcal{W}_{\rm IRB}(\theta, \phi)\Delta g^{(t)}(\theta,\phi).
\end{align}
Once the estimates of $\Delta {\rm SWE}^{(t)}_{\rm IRB}(\theta,\phi)$ and $\Delta {\rm SMS}^{(t)}_{\rm IRB}(\theta,\phi)$ is obtained, the variations in GWS over the IRB are computed as
\begin{align*}
\Delta {\rm GWS}^{(t)}_{\rm IRB}(\theta,\phi) = \Delta {\rm TWS}^{(t)}_{\rm IRB}(\theta,\phi) - \Delta {\rm SWE}^{(t)}_{\rm IRB}(\theta,\phi) \\
- \Delta {\rm SMS}^{(t)}_{\rm IRB}(\theta,\phi).
\end{align*}

\subsection{Analysis}
\begin{figure}[!t]
	\centering
	\includegraphics[scale=0.26,trim={8mm 0  0 0 }]{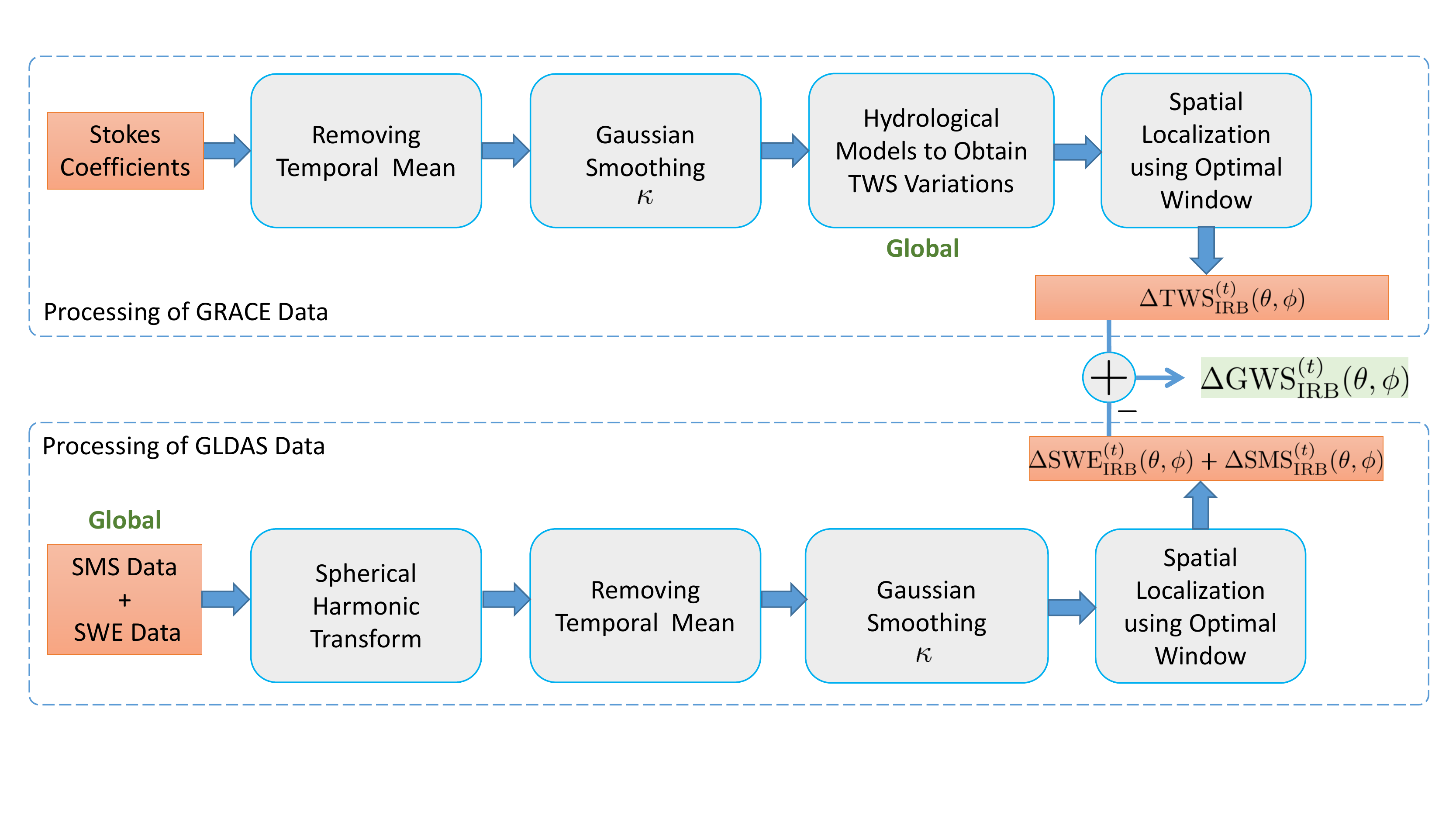}
	\vspace{-0.1in}
	\caption{Proposed Processing Framework for GRACE and GLDAS data}
	\label{fig:GRACE_block}
\end{figure}

\begin{figure}[!t]
	\centering
	%\vspace{-0.8in}
	\includegraphics[scale=.6]{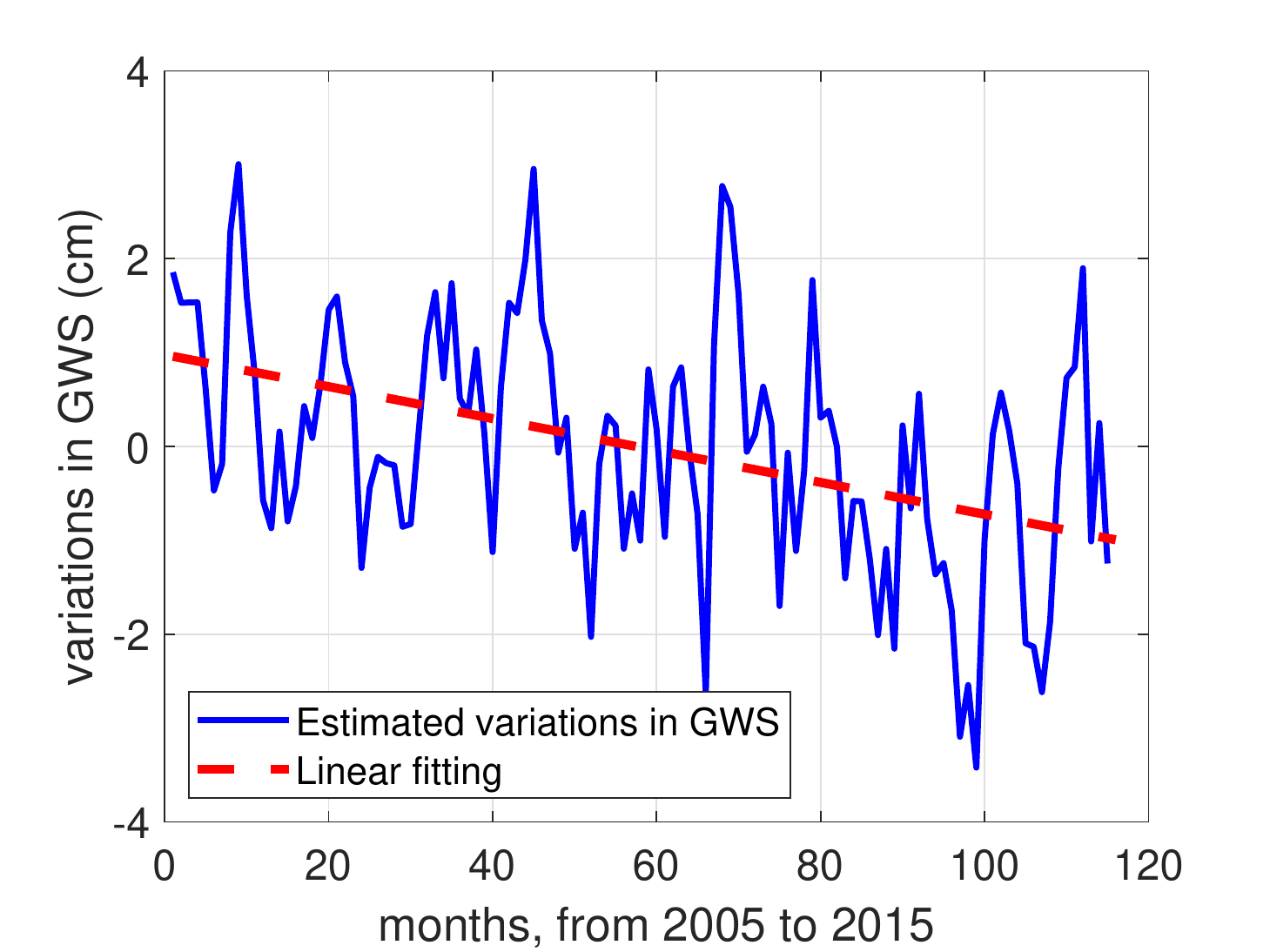}
	%\vspace{-2.5in}
	\caption{Monthly variations in GWS in the IRB from $2005$ to $2015$}
	\label{fig:GWS_var}
\end{figure}

We summarize the proposed framework to process GRACE and GLDAS data and obtain GWS variations for the region of interest in \figref{fig:GRACE_block}, which we use to process GRACE and GLDAS considering band-limit $L=61$ and IRB as our region of interest. We have used $\kappa=200$ for smoothing of the signal. We compute mean GWS variations, plotted in \figref{fig:GWS_var}, from $2005$ to $2015$. Our analysis of GWS variations reveals that it is depleting on average in IRB over time from $2005$ to $2015$.

%//////////////////////////////////////////////////////////////////////
%//////////////////////////////////////////////////////////////////////
%//////////////////////////////////////////////////////////////////////
%/////////////////////////////////////////////////////////////////////

\section{Conclusions}\label{sec:conclusion}

In this work, we have presented the data processing framework to use GRACE and GLDAS data-sets to estimate variations in TWS and GWS in both the spatial and temporal domains. We have also presented a design of optimal window function matched to the region of interest to localize these variations in the spatial domain. The proposed window function is optimal in the sense that it is band-limited in the spherical harmonic domain and has maximal energy concentration in the spatial region. Since the design of the window requires evaluation of the integral over the region of interest, we constructed optimal window for IRB by dividing the spatial region into multiple subregions and compute the integral over each subregion using analytic expressions. We have also analysed GWS variations over IRB and revealed the depletion~(on average) in GWS from $2005$ to $2015$. Our proposed processing framework can be used to analyse spatio-temporal variations in TWS and GWS for any region of interest. Potential future directions include the use of ESA's {Gravity field and steady-state Ocean Circulation Explorer~(GOCE)} mission data and consideration of significant variations in Earth’s dynamic oblateness in the data processing framework.

\bibliography{sht_bib} % needed IEEEabrv macro strings in preamble of bib file

\end{document}